\documentclass[reprint,aps,prb,twocolumn,showpacs,superscriptaddress,
amsmath,amssymb,floatfix]{revtex4-1}
\usepackage{graphicx}
\usepackage{dcolumn}
\usepackage{bm}

\begin{document}

\title{Local electronic structure of Fe$^{2+}$ impurities in MgO thin films: Temperature dependent soft x-ray absorption spectroscopy study}

\author{T. ~Haupricht}
   \affiliation{II. Physikalisches Institut, Universit{\"a}t zu K{\"o}ln,
    Z{\"u}lpicher Str. 77, 50937 K{\"o}ln, Germany}
\author{R. ~Sutarto}
    \altaffiliation[Present address: ]{Department of Physics and Astronomy, University of British Columbia, 6224 Agricultural Rd., Vancouver, British Columbia V6T 1Z1, Canada}
    \affiliation{II. Physikalisches Institut, Universit{\"a}t zu K{\"o}ln,
    Z{\"u}lpicher Str. 77, 50937 K{\"o}ln, Germany}
\author{M. W. ~Haverkort}
   \altaffiliation[Present address: ]{Max Planck Institute for Solid State Research,
    Heisenbergstr. 1, 70569 Stuttgart, Germany}
    \affiliation{II. Physikalisches Institut, Universit{\"a}t zu K{\"o}ln,
    Z{\"u}lpicher Str. 77, 50937 K{\"o}ln, Germany}
\author{H. ~Ott}
    \altaffiliation[Present address: ]{Shell Exploration \& Production B.V., Kessler Park 1, 2288 GS Rijswijk, the Netherlands}
   \affiliation{II. Physikalisches Institut, Universit{\"a}t zu K{\"o}ln,
    Z{\"u}lpicher Str. 77, 50937 K{\"o}ln, Germany}
\author{A. ~Tanaka}
   \affiliation{Department of Quantum Matter, ADSM, Hiroshima University,
    Higashi-Hiroshima 739-8530, Japan}
\author{H. H. ~Hsieh}
   \affiliation{Chung Cheng Institute of Technology, National Defense
    University, Taoyuan 335, Taiwan}
\author{H.-J. ~Lin}
   \affiliation{National Synchrotron Radiation Research Center (NSRRC),
    101 Hsin-Ann Road, Hsinchu 30077, Taiwan}
\author{C. T. ~Chen}
   \affiliation{National Synchrotron Radiation Research Center (NSRRC),
    101 Hsin-Ann Road, Hsinchu 30077, Taiwan}
\author{Z. ~Hu}
    \affiliation{Max Planck Institute for Chemical Physics of Solids,
    N\"othnitzerstr. 40, 01187 Dresden, Germany}
\author{L. H. ~Tjeng}
    \affiliation{Max Planck Institute for Chemical Physics of Solids,
    N\"othnitzerstr. 40, 01187 Dresden, Germany}

\date{\today}

\begin{abstract}

We report on the local electronic structure of Fe impurities in MgO thin films. Using soft x-ray absorption spectroscopy (XAS) we verified that the Fe impurities are all in the 2+ valence state. The fine details in the line shape of the Fe $L_{2,3}$ edges provide
direct evidence for the presence of a dynamical Jahn-Teller
distortion. We are able to determine the magnitude of the
effective $D_{4h}$ crystal field energies. We also observed a
strong temperature dependence in the spectra which we can
attribute to the thermal population of low-lying excited states
that are present due to the spin-orbit coupling in the Fe $3d$.  Using this Fe$^{2+}$ impurity system as an example, we show that an
accurate measurement of the orbital moment in Fe$_3$O$_4$ will
provide a direct estimate for the effective local low-symmetry crystal fields on the Fe$^{2+}$ sites, important for the theoretical modeling of the formation of orbital ordering.
\end{abstract}


\maketitle

Magnetite is one of the most controversially discussed systems in
solid state physics.~\cite{tsuda00} It shows a first order anomaly
in the temperature dependence of the electrical conductivity at
120~K, i.e., the famous Verwey transition~\cite{verwey39} which is
accompanied by a structural phase transition from the cubic
inverse spinel to a distorted structure. It is only very recently
that one realizes that this transition may involve not only charge
ordering of Fe$^{2+}$ and Fe$^{3+}$ ions but also $t_{2g}$ orbital
ordering at the Fe$^{2+}$
sites.\cite{wright01,wright02,schlappa08} Important in this regard are the recent results from band theory studies\cite{leonov04,Jeng04} in which the charge and orbital occupations were calculated based on
the available crystal structure data.\cite{wright01,wright02}

It is highly desired to determine experimentally the electronic
structure of Fe$_3$O$_4$ and especially the local energetics of
the Fe$^{2+}$ sites in order to test the conditions under which
the $t_{2g}$ orbital polarization and ordering can occur.
Unfortunately, a direct approach to this system is difficult since
the simultaneous presence of Fe$^{2+}$ and Fe$^{3+}$ valences as
well as octahedral and tetrahedral sites makes standard electron
spectroscopic methods to yield rather broad spectral line shapes,
i.e., too featureless for a precise analysis concerning the details
about the effective crystal fields with a symmetry lower than
$O_h$.\cite{Park97}

Here we report on our study of the electronic structure of Fe
impurities in MgO thin films. Having the local quasi-octahedral
($O_h$) symmetry and similar metal-oxygen bond lengths, the
impurity system could serve as a valuable reference for the more
complex Fe$^{2+}$ containing magnetite. Using soft x-ray
absorption spectroscopy (XAS) and an analysis based on
full multiplet cluster calculations we found that the Fe impurities are all in the 2+ charge state and that a dynamical Jahn-Teller distortion is clearly present. The spectra showed a strong temperature dependence which can be traced back to the existence of low-lying excited states due to the presence of the spin-orbit interaction. We were able to make estimates concerning the magnitude and temperature dependence of the orbital and spin contributions to
the local magnetic moments. We infer that these local effects need
to be included when interpreting the temperature dependence of the
orbital and spin moments in magnetite across the Verwey
transition.

Fe$_x$Mg$_{1-x}$O samples were prepared as polycrystalline thin
films in an ultra-high vacuum molecular beam epitaxy (MBE) system
with a base pressure of $5\times10^{-10}$~mbar. High purity Mg
and Fe metal were co-evaporated from alumina crucibles onto clean
Cu substrates. Molecular oxygen was simultaneously supplied
through a leak valve. The oxygen partial pressure was kept at
about $1\times10^{-7}$~mbar and monitored by a quadrupole mass
spectrometer during growth. The Mg effusion cell temperature was
kept at 315~$^{\circ}$C corresponding to a Mg deposition rate of
2.6~\AA/min as verified using a quartz crystal thickness monitor.
During growth the substrate temperature was kept at
250~$^{\circ}$C which led to a distillation process that allows
the growth of stoichiometric MgO. This resulted in a MgO
deposition rate of about 4~\AA/min. The thickness of the
Fe$_x$Mg$_{1-x}$O films was about 120~\AA. The use of thin films
on metallic substrates was necessary to avoid charging problems
which otherwise readily occur in electron spectroscopic
experiments on bulk insulators like MgO. Clean Cu substrates were
prepared by growing roughly 1000~\AA\ thick Cu films
\textit{in-situ} on top of atomically-flat, epi-polished and
oxygen-annealed MgO substrates.

The XAS measurements were performed at the 11A Dragon beamline of the National Synchrotron Radiation Research Center (NSRRC) in
Taiwan.\cite{chen90, chen89} The photon energy resolution at the
Fe $L_{2,3}$ edges ($h\nu$ $\approx$ 700-730~eV) was set at about
0.35~eV. The spectra were recorded using the total electron yield
(TEY) method in the normal light incidence. The base pressure of
the XAS chamber was~2~$\times$~10$^{-10}$~mbar. The MBE system was
directly connected to this XAS chamber so that the freshly
prepared samples could be transferred and measured all
\textit{in-vacuo}, thereby assuring the cleanliness and
reliability of the spectra presented here.

\begin{figure}
\includegraphics[width=.9\columnwidth]{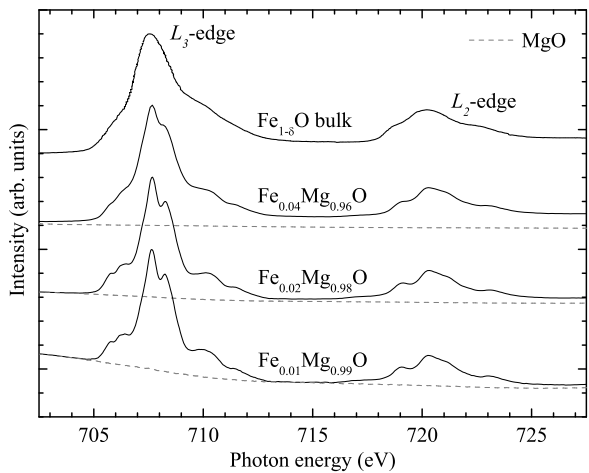}
\caption{Fe $L_{2,3}$ XAS spectra of Fe$_x$Mg$_{1-x}$O films for
\mbox{$x=0.04,0.02,0.01$}. Included is also the spectrum of bulk
Fe$_{1-\delta}$O (top curve, reproduced from Ref. \onlinecite{park}). The underlying dashed lines represent the XAS signal of a pure MgO film in the photon energy region of the Fe $L_{2,3}$, scaled to fit the pre-edge background of the respective Fe$_x$Mg$_{1-x}$O spectrum. All spectra were taken at room temperature.}\label{conc}
\end{figure}

In Fig. \ref{conc}\ we show the experimental Fe $L_{2,3}$ XAS
spectra of Fe$_x$Mg$_{1-x}$O films for $x$ = 0.04, 0.02, and 0.01.
We have also included the spectra of a bulk Fe$_{1-\delta}$O crystal
(reproduced from Ref. \onlinecite{park}) and of a pure MgO film (underlying dashed lines) as references.
The MgO spectra have been scaled to fit the pre-edge background of the respective Fe$_x$Mg$_{1-x}$O spectrum.
The pure MgO film was prepared under the same conditions as the
Fe$_x$Mg$_{1-x}$O samples, i.e., it has $x$ = 0.00. The Fe
$L_{2,3}$ spectra are dominated by the Fe $2p$ core-hole
spin-orbit coupling, which splits the spectrum roughly into two
parts, namely the $L_3$ ($h\nu \approx 708$~eV) and $L_2$ ($h\nu
\approx 721$~eV) white line regions. The line shapes of the spectra
depend strongly on the multiplet structure given by the
atomic-like Fe~$2p$-$3d$ and $3d$-$3d$ Coulomb and exchange
interactions, as well as by the surrounding solid.

In going from bulk Fe$_{1-\delta}$O to the films with decreasing
Fe concentrations, we can clearly observe that the spectral
features become sharper. This can be taken as an indication for
the presence of inter-Fe interactions in the more concentrated
systems. Here we would like to note that the presence of Fe$^{3+}$
species may also contribute to the broad spectral features of bulk
Fe$_{1-\delta}$O, a material known to have inherent
defects.\cite{jeanloz83,rao97} For Fe concentrations lower than
2\% the spectra do not significantly change anymore - apparently
here we already arrived at the impurity limit. We notice that
for the lowest Fe concentrations the pre-edge XAS background is
increasing. This can be attributed to the contribution of the MgO
to the XAS signal in the Fe $L_{2,3}$ region as shown by the
spectrum of the pure MgO film.

\begin{figure*}
\includegraphics[width=.9\textwidth]{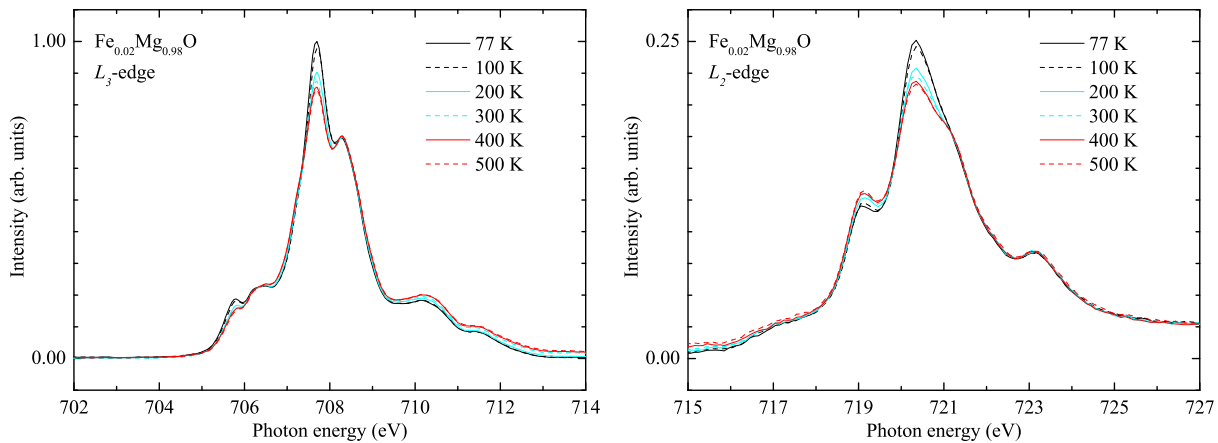}
\caption{(Color online) Temperature dependence of the experimental
Fe $L_{3}$ and $L_{2}$ XAS spectra of Fe$_{0.02}$Mg$_{0.98}$O
after subtraction of the pure MgO film background.}
\label{tdepexp}
\end{figure*}

We now focus on the temperature dependence of the spectra. Fig.
\ref{tdepexp} shows a close-up of the experimental Fe~$L_{3}$ and
$L_{2}$ XAS spectra of Fe$_{0.02}$Mg$_{0.98}$O for various
temperatures ranging from 77 up to 500~K. For clarity, we here
subtracted the XAS background coming from the pure MgO film
in the Fe $L_{2,3}$ region. Clear and systematic changes with
temperature can be observed in the spectra. This can be taken as a
direct indication for the presence of local low-lying excited
states.

To interpret and understand the spectra and their temperature
dependence, we have performed simulations of the atomic-like
$2p^{6}3d^{n} \rightarrow 2p^{5}3d^{n+1}$ ($n=6$ for Fe$^{2+}$)
transitions using the well-proven configuration-interaction
cluster model.\cite{Tanaka94,deGroot94,Thole97} Within this method we have treated the Fe impurity site as an FeO$_6$ cluster which includes the full atomic multiplet theory and the local effects of the solid.
It accounts for the intra-atomic $3d$-$3d$ and $2p$-$3d$ Coulomb
interactions, the atomic $2p$ and $3d$ spin-orbit couplings, the
local crystal field, and the O~$2p$-Fe~$3d$ hybridization. This
hybridization is taken into account by adding the
$3d^{n+1}\underline{L}$ and $3d^{n+2}\underline{L}^{2}$ etc.
states to the starting $3d^{n}$ configuration, where
$\underline{L}$ denotes a hole in the O $p$ ligands. Parameters
for the multipole part of the Coulomb interactions were given by
the Hartree-Fock values,\cite{Tanaka94} while the monopole parts
($U_{dd}$, $U_{pd}$) were estimated from photoemission experiments
on FeO.\cite{Bocquet92b} The one-electron parameters such as the O~$2p$-Fe~$3d$ charge-transfer energies and integrals as well as
the crystal field values were tuned to find the best match to the
experimental spectra. The simulations were carried out using
the program XTLS 8.3.\cite{Tanaka94,parameters}

\begin{figure}[t]
\includegraphics[width=.9\columnwidth]{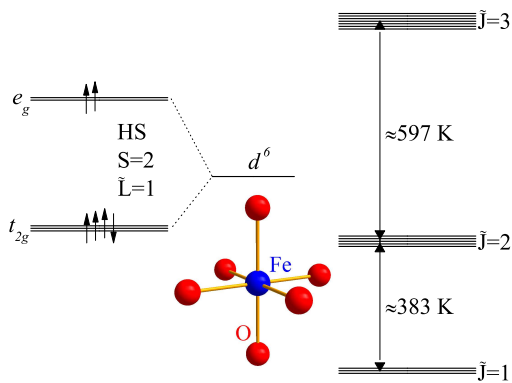}
\caption{(Color online) Energy level diagrams for a Fe$^{2+}$
cluster in the $O_h$ coordination in a crystal field (left) and a full multiplet (right) scheme. HS denotes the Hund's rule high spin configuration} \label{ediags}
\end{figure}

Starting with the simple crystal field scheme, an Fe ion in the
$O_h$ coordination will have its $3d$ states split into the lower lying $t_{2g}$ and higher $e_g$ levels, with the splitting given by 10Dq of order 1~eV (see Fig. \ref{ediags}). For an Fe$^{2+}$ ion, 5
electrons will occupy all the available spin-up states. The
remaining electron occupies one of the three spin-down $t_{2g}$ orbitals ($t_{2g}^4e_g^2$), giving the high spin $S=2$ Hund's rule ground state. In a multiplet scheme, neglecting the spin-orbit
interaction, the ground state is termed the $^{5}T_{2g}$ state,
well separated by about 1~eV or more from the higher lying
$^{5}E_{g}$ and other lower spin configurations. Assigning a
pseudo orbital momentum of $\tilde L=1$ to the open $t_{2g}$
shell,\cite{Ballhausen, Goodenough68} the spin-orbit interaction
will couple it to the $S=2$ spin, resulting in three different
states with $\tilde J=1$, 2, and 3. The presence of the
spin-orbit coupling in the $3d$ shell thus splits the 15-fold
$^{5}T_{2g}$ state into the $\tilde J$ states with degeneracies of
3, 5, and 7, respectively (see Fig. \ref{ediags}). Using typical
parameters for FeO,\cite{Tanaka94,parameters} we find an energy
separation between them of about 33~meV ($\simeq$~383~K) and 51~meV
($\simeq$~597~K), respectively.

\begin{figure}
\includegraphics[width=.9\columnwidth]{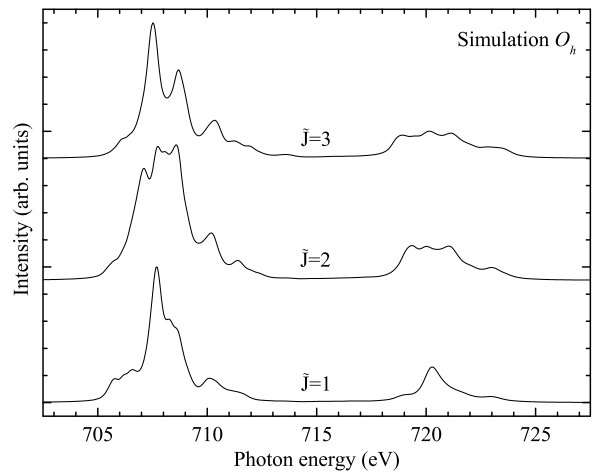}
\caption{Theoretical Fe $L_{2,3}$ XAS spectra
starting from the three lowest multiplet states of the FeO$_6$
cluster in $O_h$ symmetry, namely the $\tilde J=1$, 2, and 3
manifolds of the $^{5}T_{2g}$.} \label{Jspecs}
\end{figure}

Important for the understanding of the line shape of the Fe
$L_{2,3}$ XAS spectra and their temperature dependence is that
initial states with different quantum numbers could produce quite
different XAS spectra, since the dipole selection rules, e.g.,
$\Delta J=0,\pm1$, will dictate which of the possible final states
can be reached in the photo-absorption process. This is shown in
Fig. \ref{Jspecs}. Indeed, each of the three different $\tilde J=1$, 2, and 3 states has its own characteristic XAS spectrum. It
is then also quite natural to expect a strong temperature
dependence for the XAS spectrum of a Fe$^{2+}$ system if an
increase in temperature causes a thermal population of the $\tilde
J=2$ and 3 excited states at the expense of a depopulation of
the $\tilde J=1$ ground state.

\begin{figure}[h!]
\includegraphics[width=.9\columnwidth]{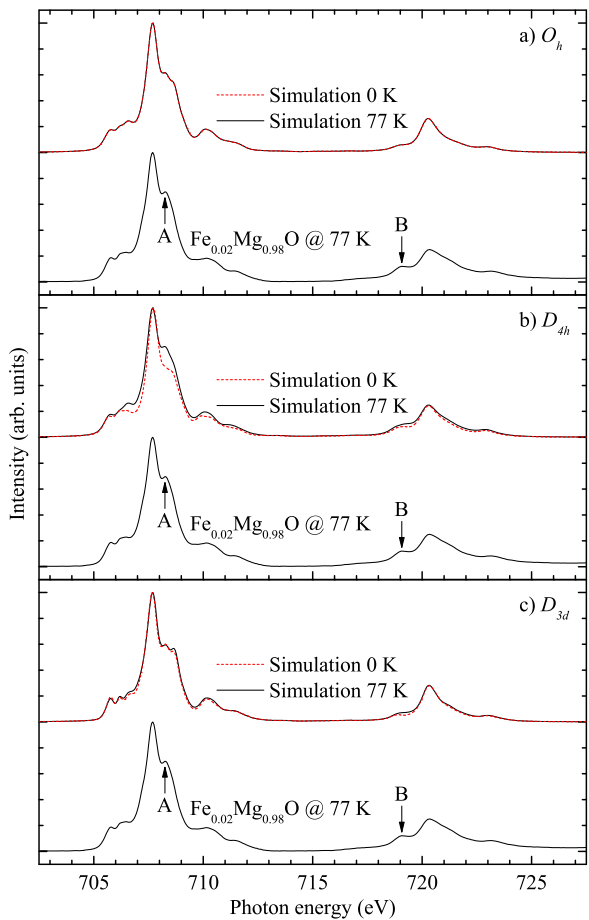}
\caption{(Color online) Comparison between the experimental 77~K
Fe $L_{2,3}$ XAS spectrum of Fe$_{0.02}$Mg$_{0.98}$O with the
simulated 0~K and 77~K spectra of the FeO$_6$ cluster in the (a)
$O_h$, (b) $D_{4h}$, and (c) $D_{3d}$ symmetry. Parameters for the
simulations are explained in the text.} \label{Comparisons}
\end{figure}

Yet, a detailed comparison between the experimental spectra and
the simulations for the $O_h$ case reveals important quantitative
discrepancies. A closer look is provided in panel (a) of Fig.
\ref{Comparisons}. One can clearly observe that neither the simulated 0~K spectrum, i.e., from the pure $\tilde J=1$ ground state, nor the simulated 77~K spectrum, i.e., containing some amount of the
$\tilde J=2$ excited state, can reproduce the experimentally
obtained 77~K spectrum. In particular, feature A is a single peak
in the experiment while the $O_h$ simulation produces two peaks,
and feature B of the experiment has considerably more weight than
can be generated by the simulation. All these strongly suggest
that the symmetry must be lower than $O_h$, in line with earlier
studies using optical and M\"{o}ssbauer
spectroscopies.\cite{Jones67,Ham67,Leider68,Manson76,Hjortsberg88}

We have investigated two further scenarios: the $D_{3d}$ (trigonal) and $D_{4h}$ (tetragonal) cases. We find that the $D_{3d}$ scenario does not provide a better fit, e.g., feature A still has a clear two peak structure and peak B has also not enough weight, as depicted in
panel (c) of Fig. \ref{Comparisons}. On the other hand, we
have been able to obtain a very good fit using the $D_{4h}$
scenario as shown in the middle panel (b) of Fig. \ref{Comparisons}:
the simulation gives more weight for peak B in better agreement
with the experiment, and the simulated peak A is a more singly peak
now as it is in the experiment. The overall line shape is thus well
reproduced. We note that this also provides evidence that the
Fe$_{0.02}$Mg$_{0.98}$O film contains only Fe$^{2+}$ ions since
the simulation has been done for an FeO$_6$ cluster having the
$3d^{n}$, $3d^{n+1}\underline{L}$, and $3d^{n+2}\underline{L}^{2}$
configurations with $n=6$. We thus find no indication for the
presence of Fe$^{3+}$ or Fe$^{1+}$ ions in our MBE-grown thin film
samples, in contrast to other earlier
studies.\cite{Leider68,Wong68,Wilkinson68,Chappert70,Manson76,Modine77}

\begin{figure*}
\includegraphics[width=.9\textwidth]{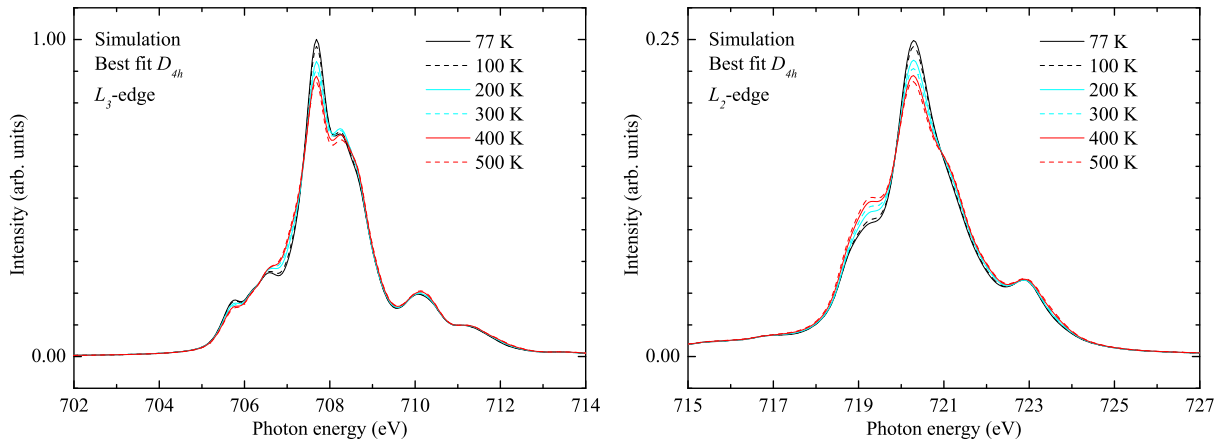}
\caption{(Color online) Simulations of the temperature dependence
of the Fe $L_{3}$ and $L_{2}$ XAS spectra in the $D_{4h}$ symmetry.
Parameters for the simulations are explained in the text.}
\label{tdeptheo}
\end{figure*}

Continuing now with the $D_{4h}$ scenario, we also have simulated
the temperature dependence of the Fe $L_{3}$ and $L_{2}$ XAS
spectra. Fig. \ref{tdeptheo} shows the results. One can observe
that the experimentally obtained temperature dependence (see Fig.
\ref{tdepexp}) is quantitatively very well reproduced. The good
agreement between simulation and experiment at all the
temperatures measured can be taken as a strong indication that the
$D_{4h}$ scenario describes accurately the local symmetry of the
Fe ion in MgO and that the model parameters chosen are realistic
giving also an appropriate energy separation between the ground
state and the excited states. In the following we will discuss in
more detail the total energy level diagram of the Fe$^{2+}$
cluster in $D_{4h}$ symmetry.

\begin{figure}
\includegraphics[width=.9\columnwidth]{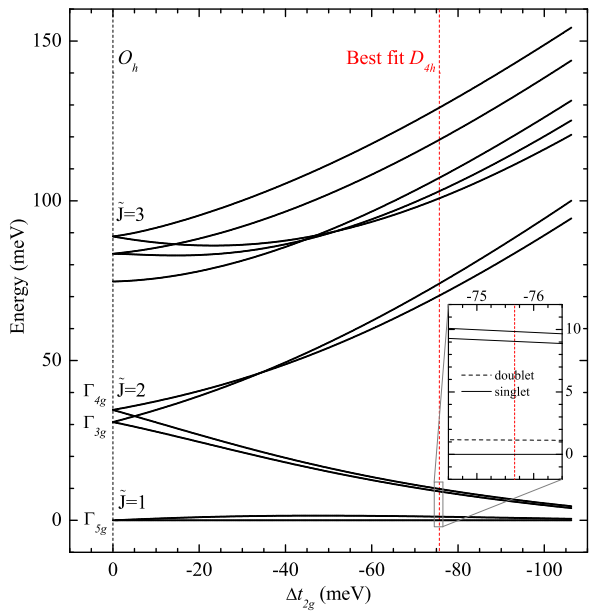}
\caption{(Color online) The total energy level diagram of the Fe$^{2+}$ FeO$_6$ cluster
as a function of the $D_{4h}$ crystal field splitting expressed in
terms of $\Delta t_{2g}$.} \label{ted}
\end{figure}

As already mentioned above, in $O_h$ symmetry the Fe $3d^{6}$
$^{5}T_{2g}$ ground state is split by the spin-orbit interaction
into the $\tilde J=1$, 2, and 3 states. A closer look reveals
that there are also smaller splittings within the $\tilde J=2$
and 3 manifolds. The ground state with $\tilde J=1$ is also
labeled as $\Gamma _{5g}$, while the higher lying first excited
states with $\tilde J=2$ are given the terms $\Gamma _{3g}$ and
$\Gamma _{4g}$. We now switch on the $D_{4h}$ crystal field by
introducing the parameters $Ds$ and $Dt$,\cite{Ballhausen} as well as differences in the O~$2p$-Fe~$3d$ hopping integrals along the
$c$-axis vs. the $a$-axis of the FeO$_{6}$ cluster. The $D_{4h}$
effective crystal field parameter can then be most conveniently
described as the effective energy splitting $\Delta t_{2g}$ between
the $xy$ and the $yz$/$zx$ states. This splitting can be
determined from a total energy calculation for which the
spin-orbit interaction is set to zero.
Here a negative $\Delta t_{2g}$ means that the $xy$ is the lowest state (compressed octahedron).
The resulting total energy diagram (including spin-orbit interaction) vs. $\Delta t_{2g}$ is
plotted in Fig. \ref{ted}.

In going from $O_h$ to $D_{4h}$ with increasing $\Delta t_{2g}$,
we find that the splitting between the $\Gamma _{5g}$ ground state
and part of the excited states $\Gamma _{3g}$ and $\Gamma _{4g}$
becomes reduced. It decreases from roughly 30~meV for $\Delta
t_{2g}=0$ to approximately 10~meV for $\Delta t_{2g}=-76$~meV,
the value with which we find the best simulations for our XAS
data. See also the inset of Fig. \ref{ted}. This 10~meV value
agrees very well with earlier optical and spin relaxation measurements which have inferred the existence of a state at about 100-115 cm$^{-1}$.\cite{Wong68,Wilkinson68} In our simulations we need to have the splitting reduced from its large cubic value of 30~meV to this particular 10~meV number in order (1) to have sufficient admixing of the $\Gamma _{3g}$ and $\Gamma _{4g}$ into the primarily $\Gamma _{5g}$-like ground state so that feature A
becomes more like a single peak and feature B gains substantial
spectral weight as in the experiment (see Fig. \ref{Comparisons}),
and (2) to obtain sufficient thermal population of the excited
states with increasing temperature in the 77-500~K range so that
the experimentally observed strong temperature dependence is
reproduced (see Figs. \ref{tdepexp} and \ref{tdeptheo}). We would
like to note that the ground state of this 3d$^{6}$ system in the
$D_{4h}$ symmetry is a singlet, as can be seen in the inset of
Fig. \ref{ted}.

As indicated above, we found $\Delta t_{2g}=-76$~meV to be the optimal number for the effective energy splitting in order to achieve the best simulations for the experimental data. It is important to
realize that this splitting lies well within the phonon energies
of the MgO crystal.\cite{Sangster70,Oganov03} Therefore, rather
than expecting to see a static Jahn-Teller distortion, as it is present in some Fe$^{2+}$ containing metal complexes, \cite{Thole88, Liao01, Bernien09} one should
take this distortion as dynamical in which the phonons are
strongly coupled to the electronic degrees of freedom as pointed
out by Ham and co-workers.\cite{Ham67,Ham69,Hjortsberg88} To
generate static distortions one would need an effective crystal
field splitting of about 0.2~eV or larger.

\begin{figure*}
\includegraphics[width=.9\textwidth]{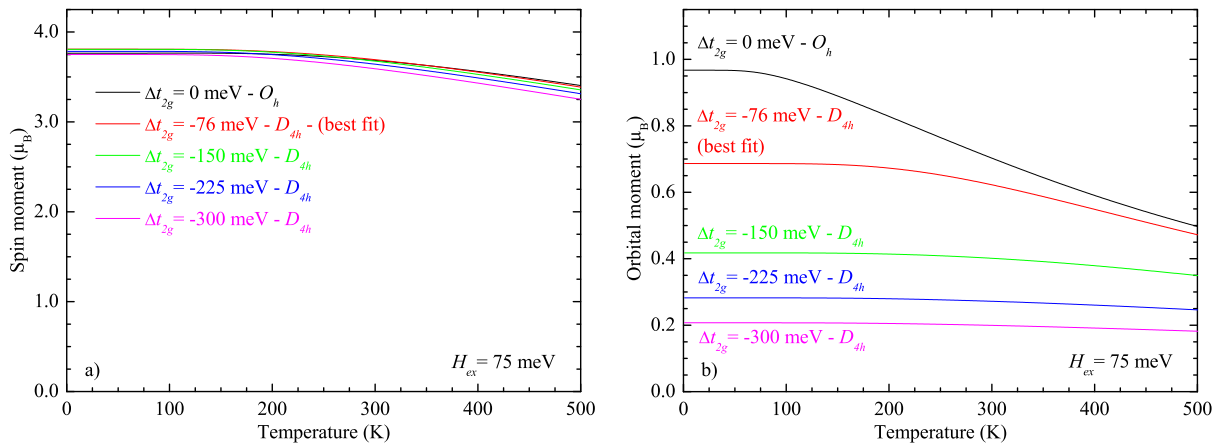}
\caption{(Color online) Simulated temperature dependence of the spin ($m_s=2\cdot S_x$, panel (a)), and orbital ($m_l=L_x$, panel (b)) contributions to the local magnetic moment of the FeO$_6$ cluster for various $D_{4h}$ crystal field energies ($\Delta t_{2g}$) and an exchange field of 75~meV.}
\label{moments}
\end{figure*}

It is now interesting to see what consequences the presence of
such a $D_{4h}$ crystal field splitting has for the magnetic
properties of the Fe$^{2+}$ ion. We calculate the spin and orbital
contributions to the magnetic moments in the presence of an
exchange field ($H_{ex}$) of 75~meV. We chose for
this value as it may be taken as a crude estimate for the case
of Fe$_3$O$_4$. Fig. \ref{moments} shows the results of the
calculations for several values of the crystal field energy
$\Delta t_{2g}$. For $\Delta t_{2g}=0$, i.e., the $O_h$ case, the
orbital moment is very large, very close to 1.0~$\mu_{B}$. Yet, it
also decreases rapidly with temperature: at 500~K it becomes 0.5~$\mu_{B}$.
This is the consequence of the thermal population of
the $\tilde J=2$ and 3 excited states. Upon switching on the
$D_{4h}$ crystal field to -76~meV, the orbital moment gets also
reduced, to about 0.7~$\mu_{B}$ already at 0~K. Increasing further
the crystal field to -150~meV, -225~meV and -300~meV produces smaller
and smaller orbital moments, i.e., about 0.4, 0.3, and 0.2~$\mu_{B}$,
respectively. The spin moment, nevertheless, always
stays close to about 4~$\mu_{B}$.

These findings could provide an interesting path to critically
test recent electronic structure theories\cite{leonov04,Jeng04}
for the explanation of the experimentally observed complex charge
and orbital order phenomena in
Fe$_3$O$_4$.\cite{wright01,wright02,schlappa08} An accurate
measurement of the orbital moment, i.e., the orbital moment at the
Fe$^{2+}$ sites (the Fe$^{3+}$ 3d$^{5}$ with their high-spin
half-filled shell do not carry an orbital moment), will provide a
direct estimate of the magnitude of the effective crystal field
splitting. This in turn will determine whether the occupied
minority $t_{2g}$ orbital of the Fe$^{2+}$ is made of mainly real
space orbitals or has a more complex nature. Only for crystal
fields substantially larger than the spin-orbit interaction one
can obtain the real space orbitals necessary to build a robust
orbital ordering. In this sense the measurement of 0.76~$\mu_{B}$
orbital moment at 88~K by Huang \textit{et al.}\cite{huang04}
would suggest the occupation of a complex $t_{2g}$ orbital and a
crystal field too small to produce a static Jahn-Teller
distortion. On the other hand, the measurement by Goering
\textit{et al.}\cite{goering06} of 0.01~$\mu_{B}$ would support
the scenario for real space orbitals and large static Jahn-Teller
distortions, much more in line with the recent theoretical
studies.\cite{leonov04,Jeng04} Nevertheless, the issue on the
magnitude of the orbital moment is not clear and is subject of
debate.\cite{goering06_comment, huang06_reply}

To conclude, we have succeeded in preparing the Fe$^{2+}$:MgO
impurity system using MBE thin film technology. The resulting Fe
$L_{2,3}$ soft x-ray absorption spectra display very sharp
features, thereby allowing us to firmly establish that the Fe
local coordination has a lower symmetry than $O_h$. Detailed
analysis of the spectral line shape and its temperature dependence
reveals that the local symmetry is $D_{4h}$ with an effective $t_{2g}$
crystal field splitting of about -76~meV. With an energy well
within the phonon frequencies of MgO, this gives rise to a dynamic
Jahn-Teller distortion. Using this Fe$^{2+}$ impurity system as a
model we showed that an accurate measurement of the orbital moment
in Fe$_3$O$_4$ will provide a direct estimate for the effective
local low-symmetry crystal fields on the Fe$^{2+}$ sites, important for the
theoretical modeling of the formation of orbital ordering.

We gratefully acknowledge the NSRRC staff for providing us with
beamtime. We would like to thank Lucie Hamdan for her skillful
technical and organizational assistance in preparing the
experiment. The research in Cologne is supported by the Deutsche
Forschungsgemeinschaft through SFB 608. T. H. is also supported by
the Bonn-Cologne Graduate School of Physics and Astronomy.


\begin{thebibliography}{39}%
\makeatletter
\providecommand \@ifxundefined [1]{%
 \@ifx{#1\undefined}
}%
\providecommand \@ifnum [1]{%
 \ifnum #1\expandafter \@firstoftwo
 \else \expandafter \@secondoftwo
 \fi
}%
\providecommand \@ifx [1]{%
 \ifx #1\expandafter \@firstoftwo
 \else \expandafter \@secondoftwo
 \fi
}%
\providecommand \natexlab [1]{#1}%
\providecommand \enquote  [1]{``#1''}%
\providecommand \bibnamefont  [1]{#1}%
\providecommand \bibfnamefont [1]{#1}%
\providecommand \citenamefont [1]{#1}%
\providecommand \href@noop [0]{\@secondoftwo}%
\providecommand \href [0]{\begingroup \@sanitize@url \@href}%
\providecommand \@href[1]{\@@startlink{#1}\@@href}%
\providecommand \@@href[1]{\endgroup#1\@@endlink}%
\providecommand \@sanitize@url [0]{\catcode `\\12\catcode `\$12\catcode
  `\&12\catcode `\#12\catcode `\^12\catcode `\_12\catcode `\%12\relax}%
\providecommand \@@startlink[1]{}%
\providecommand \@@endlink[0]{}%
\providecommand \url  [0]{\begingroup\@sanitize@url \@url }%
\providecommand \@url [1]{\endgroup\@href {#1}{\urlprefix }}%
\providecommand \urlprefix  [0]{URL }%
\providecommand \Eprint [0]{\href }%
\@ifxundefined \urlstyle {%
  \providecommand \doi  [0]{\begingroup \@sanitize@url \@doi}%
  \providecommand \@doi [1]{\endgroup \@@startlink {\doibase
  #1}doi:\discretionary {}{}{}#1\@@endlink }%
}{%
  \providecommand \doi  [0]{doi:\discretionary{}{}{}\begingroup
  \urlstyle{rm}\Url }%
}%
\providecommand \doibase [0]{http://dx.doi.org/}%
\providecommand \Doi [0]{\begingroup \@sanitize@url \@Doi }%
\providecommand \@Doi  [1]{\endgroup\@@startlink{\doibase#1}\@@Doi}%
\providecommand \@@Doi [1]{#1\@@endlink}%
\providecommand \selectlanguage [0]{\@gobble}%
\providecommand \bibinfo  [0]{\@secondoftwo}%
\providecommand \bibfield  [0]{\@secondoftwo}%
\providecommand \translation [1]{[#1]}%
\providecommand \BibitemOpen [0]{}%
\providecommand \bibitemStop [0]{}%
\providecommand \bibitemNoStop [0]{.\EOS\space}%
\providecommand \EOS [0]{\spacefactor3000\relax}%
\providecommand \BibitemShut  [1]{\csname bibitem#1\endcsname}%
\bibitem [{\citenamefont {Tsuda}\ \emph {et~al.}(2000)\citenamefont {Tsuda},
  \citenamefont {Nasu}, \citenamefont {Fujimori},\ and\ \citenamefont
  {Siratori}}]{tsuda00}%
  \BibitemOpen
  \bibfield  {author} {\bibinfo {author} {\bibfnamefont {N.}~\bibnamefont
  {Tsuda}}, \bibinfo {author} {\bibfnamefont {K.}~\bibnamefont {Nasu}},
  \bibinfo {author} {\bibfnamefont {A.}~\bibnamefont {Fujimori}}, \ and\
  \bibinfo {author} {\bibfnamefont {K.}~\bibnamefont {Siratori}},\ }\href@noop
  {} {\emph {\bibinfo {title} {Electronic Conduction in Oxides}}}\ (\bibinfo
  {publisher} {Springer, New York},\ \bibinfo {year} {2000})\BibitemShut
  {NoStop}%
\bibitem [{\citenamefont {Verwey}(1939)}]{verwey39}%
  \BibitemOpen
  \bibfield  {author} {\bibinfo {author} {\bibfnamefont {E.~J.~W.}\
  \bibnamefont {Verwey}},\ }\Doi {10.1038/144327b0} {\bibfield  {journal}
  {\bibinfo  {journal} {Nature},\ }\textbf {\bibinfo {volume} {144}},\ \bibinfo
  {pages} {327} (\bibinfo {year} {1939})}\BibitemShut {NoStop}%
\bibitem [{\citenamefont {Wright}\ \emph {et~al.}(2001)\citenamefont {Wright},
  \citenamefont {Attfield},\ and\ \citenamefont {Radaelli}}]{wright01}%
  \BibitemOpen
  \bibfield  {author} {\bibinfo {author} {\bibfnamefont {J.~P.}\ \bibnamefont
  {Wright}}, \bibinfo {author} {\bibfnamefont {J.~P.}\ \bibnamefont
  {Attfield}}, \ and\ \bibinfo {author} {\bibfnamefont {P.~G.}\ \bibnamefont
  {Radaelli}},\ }\Doi {10.1103/PhysRevLett.87.266401} {\bibfield  {journal}
  {\bibinfo  {journal} {Phys. Rev. Lett.},\ }\textbf {\bibinfo {volume} {87}},\
  \bibinfo {pages} {266401} (\bibinfo {year} {2001})}\BibitemShut {NoStop}%
\bibitem [{\citenamefont {Wright}\ \emph {et~al.}(2002)\citenamefont {Wright},
  \citenamefont {Attfield},\ and\ \citenamefont {Radaelli}}]{wright02}%
  \BibitemOpen
  \bibfield  {author} {\bibinfo {author} {\bibfnamefont {J.~P.}\ \bibnamefont
  {Wright}}, \bibinfo {author} {\bibfnamefont {J.~P.}\ \bibnamefont
  {Attfield}}, \ and\ \bibinfo {author} {\bibfnamefont {P.~G.}\ \bibnamefont
  {Radaelli}},\ }\Doi {10.1103/PhysRevB.66.214422} {\bibfield  {journal}
  {\bibinfo  {journal} {Phys. Rev. B},\ }\textbf {\bibinfo {volume} {66}},\
  \bibinfo {pages} {214422} (\bibinfo {year} {2002})}\BibitemShut {NoStop}%
\bibitem [{\citenamefont {Schlappa}\ \emph {et~al.}(2008)\citenamefont
  {Schlappa}, \citenamefont {{Sch\"{u}\ss ler-Langeheine}}, \citenamefont
  {Chang}, \citenamefont {Ott}, \citenamefont {Tanaka}, \citenamefont {Hu},
  \citenamefont {Haverkort}, \citenamefont {Schierle}, \citenamefont {Weschke},
  \citenamefont {Kaindl},\ and\ \citenamefont {Tjeng}}]{schlappa08}%
  \BibitemOpen
  \bibfield  {author} {\bibinfo {author} {\bibfnamefont {J.}~\bibnamefont
  {Schlappa}}, \bibinfo {author} {\bibfnamefont {C.}~\bibnamefont {{Sch\"{u}\ss
  ler-Langeheine}}}, \bibinfo {author} {\bibfnamefont {C.~F.}\ \bibnamefont
  {Chang}}, \bibinfo {author} {\bibfnamefont {H.}~\bibnamefont {Ott}}, \bibinfo
  {author} {\bibfnamefont {A.}~\bibnamefont {Tanaka}}, \bibinfo {author}
  {\bibfnamefont {Z.}~\bibnamefont {Hu}}, \bibinfo {author} {\bibfnamefont
  {M.~W.}\ \bibnamefont {Haverkort}}, \bibinfo {author} {\bibfnamefont
  {E.}~\bibnamefont {Schierle}}, \bibinfo {author} {\bibfnamefont
  {E.}~\bibnamefont {Weschke}}, \bibinfo {author} {\bibfnamefont
  {G.}~\bibnamefont {Kaindl}}, \ and\ \bibinfo {author} {\bibfnamefont {L.~H.}\
  \bibnamefont {Tjeng}},\ }\Doi {10.1103/PhysRevLett.100.026406} {\bibfield
  {journal} {\bibinfo  {journal} {Phys. Rev. Lett.},\ }\textbf {\bibinfo
  {volume} {100}},\ \bibinfo {eid} {026406} (\bibinfo {year}
  {2008})}\BibitemShut {NoStop}%
\bibitem [{\citenamefont {Leonov}\ \emph {et~al.}(2004)\citenamefont {Leonov},
  \citenamefont {Yaresko}, \citenamefont {Antonov}, \citenamefont {Korotin},\
  and\ \citenamefont {Anisimov}}]{leonov04}%
  \BibitemOpen
  \bibfield  {author} {\bibinfo {author} {\bibfnamefont {I.}~\bibnamefont
  {Leonov}}, \bibinfo {author} {\bibfnamefont {A.~N.}\ \bibnamefont {Yaresko}},
  \bibinfo {author} {\bibfnamefont {V.~N.}\ \bibnamefont {Antonov}}, \bibinfo
  {author} {\bibfnamefont {M.~A.}\ \bibnamefont {Korotin}}, \ and\ \bibinfo
  {author} {\bibfnamefont {V.~I.}\ \bibnamefont {Anisimov}},\ }\Doi
  {10.1103/PhysRevLett.93.146404} {\bibfield  {journal} {\bibinfo  {journal}
  {Phys. Rev. Lett.},\ }\textbf {\bibinfo {volume} {93}},\ \bibinfo {pages}
  {146404} (\bibinfo {year} {2004})}\BibitemShut {NoStop}%
\bibitem [{\citenamefont {Jeng}\ \emph {et~al.}(2004)\citenamefont {Jeng},
  \citenamefont {Guo},\ and\ \citenamefont {Huang}}]{Jeng04}%
  \BibitemOpen
  \bibfield  {author} {\bibinfo {author} {\bibfnamefont {H.-T.}\ \bibnamefont
  {Jeng}}, \bibinfo {author} {\bibfnamefont {G.~Y.}\ \bibnamefont {Guo}}, \
  and\ \bibinfo {author} {\bibfnamefont {D.~J.}\ \bibnamefont {Huang}},\ }\Doi
  {10.1103/PhysRevLett.93.156403} {\bibfield  {journal} {\bibinfo  {journal}
  {Phys. Rev. Lett.},\ }\textbf {\bibinfo {volume} {93}},\ \bibinfo {pages}
  {156403} (\bibinfo {year} {2004})}\BibitemShut {NoStop}%
\bibitem [{\citenamefont {Park}\ \emph {et~al.}(1997)\citenamefont {Park},
  \citenamefont {Tjeng}, \citenamefont {Allen}, \citenamefont {Metcalf},\ and\
  \citenamefont {Chen}}]{Park97}%
  \BibitemOpen
  \bibfield  {author} {\bibinfo {author} {\bibfnamefont {J.-H.}\ \bibnamefont
  {Park}}, \bibinfo {author} {\bibfnamefont {L.~H.}\ \bibnamefont {Tjeng}},
  \bibinfo {author} {\bibfnamefont {J.~W.}\ \bibnamefont {Allen}}, \bibinfo
  {author} {\bibfnamefont {P.}~\bibnamefont {Metcalf}}, \ and\ \bibinfo
  {author} {\bibfnamefont {C.~T.}\ \bibnamefont {Chen}},\ }\Doi
  {10.1103/PhysRevB.55.12813} {\bibfield  {journal} {\bibinfo  {journal} {Phys.
  Rev. B},\ }\textbf {\bibinfo {volume} {55}},\ \bibinfo {pages} {12813}
  (\bibinfo {year} {1997})}\BibitemShut {NoStop}%
\bibitem [{\citenamefont {Chen}\ and\ \citenamefont {Sette}(1990)}]{chen90}%
  \BibitemOpen
  \bibfield  {author} {\bibinfo {author} {\bibfnamefont {C.~T.}\ \bibnamefont
  {Chen}}\ and\ \bibinfo {author} {\bibfnamefont {F.}~\bibnamefont {Sette}},\
  }\Doi {10.1088/0031-8949/1990/T31/016} {\bibfield  {journal} {\bibinfo
  {journal} {Phys. Scr.},\ }\textbf {\bibinfo {volume} {T31}},\ \bibinfo
  {pages} {119} (\bibinfo {year} {1990})}\BibitemShut {NoStop}%
\bibitem [{\citenamefont {Chen}\ and\ \citenamefont {Sette}(1989)}]{chen89}%
  \BibitemOpen
  \bibfield  {author} {\bibinfo {author} {\bibfnamefont {C.~T.}\ \bibnamefont
  {Chen}}\ and\ \bibinfo {author} {\bibfnamefont {F.}~\bibnamefont {Sette}},\
  }\Doi {10.1063/1.1141044} {\bibfield  {journal} {\bibinfo  {journal} {Rev.
  Sci. Instrum.},\ }\textbf {\bibinfo {volume} {60}},\ \bibinfo {pages} {1616}
  (\bibinfo {year} {1989})}\BibitemShut {NoStop}%
\bibitem [{\citenamefont {Park}(1994)}]{park}%
  \BibitemOpen
  \bibfield  {author} {\bibinfo {author} {\bibfnamefont {J.-H.}\ \bibnamefont
  {Park}},\ }\href@noop {} {Ph.D. thesis},\ \bibinfo  {school} {University of
  Michigan} (\bibinfo {year} {1994})\BibitemShut {NoStop}%
\bibitem [{\citenamefont {Jeanloz}\ and\ \citenamefont
  {Hazen}(1983)}]{jeanloz83}%
  \BibitemOpen
  \bibfield  {author} {\bibinfo {author} {\bibfnamefont {R.}~\bibnamefont
  {Jeanloz}}\ and\ \bibinfo {author} {\bibfnamefont {R.~M.}\ \bibnamefont
  {Hazen}},\ }\Doi {10.1038/304620a0} {\bibfield  {journal} {\bibinfo
  {journal} {Nature},\ }\textbf {\bibinfo {volume} {304}},\ \bibinfo {pages}
  {620} (\bibinfo {year} {1983})}\BibitemShut {NoStop}%
\bibitem [{\citenamefont {Rao}\ and\ \citenamefont {Raveau}(1997)}]{rao97}%
  \BibitemOpen
  \bibfield  {author} {\bibinfo {author} {\bibfnamefont {C.~N.~R.}\
  \bibnamefont {Rao}}\ and\ \bibinfo {author} {\bibfnamefont {B.}~\bibnamefont
  {Raveau}},\ }\href@noop {} {\emph {\bibinfo {title} {Transition Metal
  Oxides}}},\ \bibinfo {edition} {2nd}\ ed.\ (\bibinfo  {publisher}
  {Wiley-VCH},\ \bibinfo {year} {1997})\BibitemShut {NoStop}%
\bibitem [{\citenamefont {Tanaka}\ and\ \citenamefont {Jo}(1994)}]{Tanaka94}%
  \BibitemOpen
  \bibfield  {author} {\bibinfo {author} {\bibfnamefont {A.}~\bibnamefont
  {Tanaka}}\ and\ \bibinfo {author} {\bibfnamefont {T.}~\bibnamefont {Jo}},\
  }\Doi {10.1143/JPSJ.63.2788} {\bibfield  {journal} {\bibinfo  {journal} {J.
  Phys. Soc. Jpn.},\ }\textbf {\bibinfo {volume} {63}},\ \bibinfo {pages}
  {2788} (\bibinfo {year} {1994})}\BibitemShut {NoStop}%
\bibitem [{\citenamefont {{See review by F. M. F. de
  Groot}}(1994)}]{deGroot94}%
  \BibitemOpen
  \bibfield  {author} {\bibinfo {author} {\bibnamefont {{See review by F. M. F.
  de Groot}}},\ }\Doi {10.1016/0368-2048(93)02041-J} {\bibfield  {journal}
  {\bibinfo  {journal} {J. Electron. Spectrosc. Relat. Phenom.},\ }\textbf
  {\bibinfo {volume} {67}},\ \bibinfo {pages} {529} (\bibinfo {year}
  {1994})}\BibitemShut {NoStop}%
\bibitem [{\citenamefont {{See review in the Theo Thole Memorial
  Issue}}(1997)}]{Thole97}%
  \BibitemOpen
  \bibfield  {author} {\bibinfo {author} {\bibnamefont {{See review in the Theo
  Thole Memorial Issue}}},\ }\Doi {10.1016/S0368-2048(97)00039-X} {\bibfield
  {journal} {\bibinfo  {journal} {J. Electron. Spectrosc. Relat. Phenom.},\
  }\textbf {\bibinfo {volume} {86}},\ \bibinfo {pages} {1} (\bibinfo {year}
  {1997})}\BibitemShut {NoStop}%
\bibitem [{\citenamefont {Bocquet}\ \emph {et~al.}(1992)\citenamefont
  {Bocquet}, \citenamefont {Mizokawa}, \citenamefont {Saitoh}, \citenamefont
  {Namatame},\ and\ \citenamefont {Fujimori}}]{Bocquet92b}%
  \BibitemOpen
  \bibfield  {author} {\bibinfo {author} {\bibfnamefont {A.~E.}\ \bibnamefont
  {Bocquet}}, \bibinfo {author} {\bibfnamefont {T.}~\bibnamefont {Mizokawa}},
  \bibinfo {author} {\bibfnamefont {T.}~\bibnamefont {Saitoh}}, \bibinfo
  {author} {\bibfnamefont {H.}~\bibnamefont {Namatame}}, \ and\ \bibinfo
  {author} {\bibfnamefont {A.}~\bibnamefont {Fujimori}},\ }\Doi
  {10.1103/PhysRevB.46.3771} {\bibfield  {journal} {\bibinfo  {journal} {Phys.
  Rev. B},\ }\textbf {\bibinfo {volume} {46}},\ \bibinfo {pages} {3771}
  (\bibinfo {year} {1992})}\BibitemShut {NoStop}%
\bibitem [{par()}]{parameters}%
  \BibitemOpen
  \href@noop {} {}\bibinfo {note} {{Parameters used for the calculation of the
  FeO$_6$ cluster (in eV): $\Delta$=7.5, $U_{dd}$=6.0, $U_{\underline cd}$=7.5,
  $10Dq$=0.6, $T_{pp}$=0.7, and $\zeta$ see Ref. \onlinecite{Tanaka94}, Slater
  integrals 75\% of Hartree-Fock values}}\BibitemShut {NoStop}%
\bibitem [{\citenamefont {Ballhausen}(1962)}]{Ballhausen}%
  \BibitemOpen
  \bibfield  {author} {\bibinfo {author} {\bibfnamefont {C.~J.}\ \bibnamefont
  {Ballhausen}},\ }\href@noop {} {\emph {\bibinfo {title} {Introduction to
  ligand field theory}}}\ (\bibinfo  {publisher} {McGraw-Hill},\ \bibinfo
  {address} {New York},\ \bibinfo {year} {1962})\BibitemShut {NoStop}%
\bibitem [{\citenamefont {Goodenough}(1968)}]{Goodenough68}%
  \BibitemOpen
  \bibfield  {author} {\bibinfo {author} {\bibfnamefont {J.~B.}\ \bibnamefont
  {Goodenough}},\ }\Doi {10.1103/PhysRev.171.466} {\bibfield  {journal}
  {\bibinfo  {journal} {Phys. Rev.},\ }\textbf {\bibinfo {volume} {171}},\
  \bibinfo {pages} {466} (\bibinfo {year} {1968})}\BibitemShut {NoStop}%
\bibitem [{\citenamefont {Jones}(1967)}]{Jones67}%
  \BibitemOpen
  \bibfield  {author} {\bibinfo {author} {\bibfnamefont {G.~D.}\ \bibnamefont
  {Jones}},\ }\Doi {10.1103/PhysRev.155.259} {\bibfield  {journal} {\bibinfo
  {journal} {Phys. Rev.},\ }\textbf {\bibinfo {volume} {155}},\ \bibinfo
  {pages} {259} (\bibinfo {year} {1967})}\BibitemShut {NoStop}%
\bibitem [{\citenamefont {Ham}(1967)}]{Ham67}%
  \BibitemOpen
  \bibfield  {author} {\bibinfo {author} {\bibfnamefont {F.~S.}\ \bibnamefont
  {Ham}},\ }\Doi {10.1103/PhysRev.160.328} {\bibfield  {journal} {\bibinfo
  {journal} {Phys. Rev.},\ }\textbf {\bibinfo {volume} {160}},\ \bibinfo
  {pages} {328} (\bibinfo {year} {1967})}\BibitemShut {NoStop}%
\bibitem [{\citenamefont {Leider}\ and\ \citenamefont
  {Pipkorn}(1968)}]{Leider68}%
  \BibitemOpen
  \bibfield  {author} {\bibinfo {author} {\bibfnamefont {H.~R.}\ \bibnamefont
  {Leider}}\ and\ \bibinfo {author} {\bibfnamefont {D.~N.}\ \bibnamefont
  {Pipkorn}},\ }\Doi {10.1103/PhysRev.165.494} {\bibfield  {journal} {\bibinfo
  {journal} {Phys. Rev.},\ }\textbf {\bibinfo {volume} {165}},\ \bibinfo
  {pages} {494} (\bibinfo {year} {1968})}\BibitemShut {NoStop}%
\bibitem [{\citenamefont {Manson}\ \emph {et~al.}(1976)\citenamefont {Manson},
  \citenamefont {Gourley}, \citenamefont {Vance}, \citenamefont {Sengupta},\
  and\ \citenamefont {Smith}}]{Manson76}%
  \BibitemOpen
  \bibfield  {author} {\bibinfo {author} {\bibfnamefont {N.~B.}\ \bibnamefont
  {Manson}}, \bibinfo {author} {\bibfnamefont {J.~T.}\ \bibnamefont {Gourley}},
  \bibinfo {author} {\bibfnamefont {E.~R.}\ \bibnamefont {Vance}}, \bibinfo
  {author} {\bibfnamefont {D.}~\bibnamefont {Sengupta}}, \ and\ \bibinfo
  {author} {\bibfnamefont {G.}~\bibnamefont {Smith}},\ }\Doi
  {10.1016/0022-3697(76)90144-X} {\bibfield  {journal} {\bibinfo  {journal} {J.
  Phys. Chem. Solids},\ }\textbf {\bibinfo {volume} {37}},\ \bibinfo {pages}
  {1145 } (\bibinfo {year} {1976})}\BibitemShut {NoStop}%
\bibitem [{\citenamefont {Hjortsberg}\ \emph {et~al.}(1988)\citenamefont
  {Hjortsberg}, \citenamefont {Vallin},\ and\ \citenamefont
  {Ham}}]{Hjortsberg88}%
  \BibitemOpen
  \bibfield  {author} {\bibinfo {author} {\bibfnamefont {A.}~\bibnamefont
  {Hjortsberg}}, \bibinfo {author} {\bibfnamefont {J.~T.}\ \bibnamefont
  {Vallin}}, \ and\ \bibinfo {author} {\bibfnamefont {F.~S.}\ \bibnamefont
  {Ham}},\ }\Doi {10.1103/PhysRevB.37.3196} {\bibfield  {journal} {\bibinfo
  {journal} {Phys. Rev. B},\ }\textbf {\bibinfo {volume} {37}},\ \bibinfo
  {pages} {3196} (\bibinfo {year} {1988})}\BibitemShut {NoStop}%
\bibitem [{\citenamefont {Wong}(1968)}]{Wong68}%
  \BibitemOpen
  \bibfield  {author} {\bibinfo {author} {\bibfnamefont {J.~Y.}\ \bibnamefont
  {Wong}},\ }\Doi {10.1103/PhysRev.168.337} {\bibfield  {journal} {\bibinfo
  {journal} {Phys. Rev.},\ }\textbf {\bibinfo {volume} {168}},\ \bibinfo
  {pages} {337} (\bibinfo {year} {1968})}\BibitemShut {NoStop}%
\bibitem [{\citenamefont {Wilkinson}\ \emph {et~al.}(1968)\citenamefont
  {Wilkinson}, \citenamefont {Hartman},\ and\ \citenamefont
  {Castle}}]{Wilkinson68}%
  \BibitemOpen
  \bibfield  {author} {\bibinfo {author} {\bibfnamefont {E.~L.}\ \bibnamefont
  {Wilkinson}}, \bibinfo {author} {\bibfnamefont {R.~L.}\ \bibnamefont
  {Hartman}}, \ and\ \bibinfo {author} {\bibfnamefont {J.~G.}\ \bibnamefont
  {Castle}},\ }\Doi {10.1103/PhysRev.171.299} {\bibfield  {journal} {\bibinfo
  {journal} {Phys. Rev.},\ }\textbf {\bibinfo {volume} {171}},\ \bibinfo
  {pages} {299} (\bibinfo {year} {1968})}\BibitemShut {NoStop}%
\bibitem [{\citenamefont {Chappert}\ \emph {et~al.}(1970)\citenamefont
  {Chappert}, \citenamefont {Misetich}, \citenamefont {Frankel},\ and\
  \citenamefont {Blum}}]{Chappert70}%
  \BibitemOpen
  \bibfield  {author} {\bibinfo {author} {\bibfnamefont {J.}~\bibnamefont
  {Chappert}}, \bibinfo {author} {\bibfnamefont {A.}~\bibnamefont {Misetich}},
  \bibinfo {author} {\bibfnamefont {R.~B.}\ \bibnamefont {Frankel}}, \ and\
  \bibinfo {author} {\bibfnamefont {N.~A.}\ \bibnamefont {Blum}},\ }\Doi
  {10.1103/PhysRevB.1.1929} {\bibfield  {journal} {\bibinfo  {journal} {Phys.
  Rev. B},\ }\textbf {\bibinfo {volume} {1}},\ \bibinfo {pages} {1929}
  (\bibinfo {year} {1970})}\BibitemShut {NoStop}%
\bibitem [{\citenamefont {Modine}\ \emph {et~al.}(1977)\citenamefont {Modine},
  \citenamefont {Sonder},\ and\ \citenamefont {Weeks}}]{Modine77}%
  \BibitemOpen
  \bibfield  {author} {\bibinfo {author} {\bibfnamefont {F.~A.}\ \bibnamefont
  {Modine}}, \bibinfo {author} {\bibfnamefont {E.}~\bibnamefont {Sonder}}, \
  and\ \bibinfo {author} {\bibfnamefont {R.~A.}\ \bibnamefont {Weeks}},\ }\Doi
  {10.1063/1.324201} {\bibfield  {journal} {\bibinfo  {journal} {J. Appl.
  Phys.},\ }\textbf {\bibinfo {volume} {48}},\ \bibinfo {pages} {3514}
  (\bibinfo {year} {1977})}\BibitemShut {NoStop}%
\bibitem [{\citenamefont {Sangster}\ \emph {et~al.}(1970)\citenamefont
  {Sangster}, \citenamefont {Peckham},\ and\ \citenamefont
  {Saunderson}}]{Sangster70}%
  \BibitemOpen
  \bibfield  {author} {\bibinfo {author} {\bibfnamefont {M.~J.~L.}\
  \bibnamefont {Sangster}}, \bibinfo {author} {\bibfnamefont {G.}~\bibnamefont
  {Peckham}}, \ and\ \bibinfo {author} {\bibfnamefont {D.~H.}\ \bibnamefont
  {Saunderson}},\ }\Doi {10.1088/0022-3719/3/5/017} {\bibfield  {journal}
  {\bibinfo  {journal} {J. Phys. C: Solid State Phys.},\ }\textbf {\bibinfo
  {volume} {3}},\ \bibinfo {pages} {1026} (\bibinfo {year} {1970})}\BibitemShut
  {NoStop}%
\bibitem [{\citenamefont {Oganov}\ \emph {et~al.}(2003)\citenamefont {Oganov},
  \citenamefont {Gillan},\ and\ \citenamefont {Price}}]{Oganov03}%
  \BibitemOpen
  \bibfield  {author} {\bibinfo {author} {\bibfnamefont {A.~R.}\ \bibnamefont
  {Oganov}}, \bibinfo {author} {\bibfnamefont {M.~J.}\ \bibnamefont {Gillan}},
  \ and\ \bibinfo {author} {\bibfnamefont {G.~D.}\ \bibnamefont {Price}},\
  }\Doi {10.1063/1.1570394} {\bibfield  {journal} {\bibinfo  {journal} {J.
  Chem. Phys.},\ }\textbf {\bibinfo {volume} {118}},\ \bibinfo {pages} {10174}
  (\bibinfo {year} {2003})}\BibitemShut {NoStop}%
\bibitem [{\citenamefont {Thole}\ \emph {et~al.}(1988)\citenamefont {Thole},
  \citenamefont {{van der Laan}},\ and\ \citenamefont {Butler}}]{Thole88}%
  \BibitemOpen
  \bibfield  {author} {\bibinfo {author} {\bibfnamefont {B.}~\bibnamefont
  {Thole}}, \bibinfo {author} {\bibfnamefont {G.}~\bibnamefont {{van der
  Laan}}}, \ and\ \bibinfo {author} {\bibfnamefont {P.}~\bibnamefont
  {Butler}},\ }\Doi {10.1016/0009-2614(88)85029-2} {\bibfield  {journal}
  {\bibinfo  {journal} {Chem. Phys. Lett.},\ }\textbf {\bibinfo {volume}
  {149}},\ \bibinfo {pages} {295} (\bibinfo {year} {1988})}\BibitemShut
  {NoStop}%
\bibitem [{\citenamefont {Liao}\ and\ \citenamefont {Scheiner}(2001)}]{Liao01}%
  \BibitemOpen
  \bibfield  {author} {\bibinfo {author} {\bibfnamefont {M.-S.}\ \bibnamefont
  {Liao}}\ and\ \bibinfo {author} {\bibfnamefont {S.}~\bibnamefont
  {Scheiner}},\ }\Doi {10.1063/1.1367374} {\bibfield  {journal} {\bibinfo
  {journal} {J. Chem. Phys.},\ }\textbf {\bibinfo {volume} {114}},\ \bibinfo
  {pages} {9780} (\bibinfo {year} {2001})}\BibitemShut {NoStop}%
\bibitem [{\citenamefont {Bernien}\ \emph {et~al.}(2009)\citenamefont
  {Bernien}, \citenamefont {Miguel}, \citenamefont {Weis}, \citenamefont {Ali},
  \citenamefont {Kurde}, \citenamefont {Krumme}, \citenamefont {Panchmatia},
  \citenamefont {Sanyal}, \citenamefont {Piantek}, \citenamefont {Srivastava},
  \citenamefont {Baberschke}, \citenamefont {Oppeneer}, \citenamefont
  {Eriksson}, \citenamefont {Kuch},\ and\ \citenamefont {Wende}}]{Bernien09}%
  \BibitemOpen
  \bibfield  {author} {\bibinfo {author} {\bibfnamefont {M.}~\bibnamefont
  {Bernien}}, \bibinfo {author} {\bibfnamefont {J.}~\bibnamefont {Miguel}},
  \bibinfo {author} {\bibfnamefont {C.}~\bibnamefont {Weis}}, \bibinfo {author}
  {\bibfnamefont {M.~E.}\ \bibnamefont {Ali}}, \bibinfo {author} {\bibfnamefont
  {J.}~\bibnamefont {Kurde}}, \bibinfo {author} {\bibfnamefont
  {B.}~\bibnamefont {Krumme}}, \bibinfo {author} {\bibfnamefont {P.~M.}\
  \bibnamefont {Panchmatia}}, \bibinfo {author} {\bibfnamefont
  {B.}~\bibnamefont {Sanyal}}, \bibinfo {author} {\bibfnamefont
  {M.}~\bibnamefont {Piantek}}, \bibinfo {author} {\bibfnamefont
  {P.}~\bibnamefont {Srivastava}}, \bibinfo {author} {\bibfnamefont
  {K.}~\bibnamefont {Baberschke}}, \bibinfo {author} {\bibfnamefont {P.~M.}\
  \bibnamefont {Oppeneer}}, \bibinfo {author} {\bibfnamefont {O.}~\bibnamefont
  {Eriksson}}, \bibinfo {author} {\bibfnamefont {W.}~\bibnamefont {Kuch}}, \
  and\ \bibinfo {author} {\bibfnamefont {H.}~\bibnamefont {Wende}},\ }\Doi
  {10.1103/PhysRevLett.102.047202} {\bibfield  {journal} {\bibinfo  {journal}
  {Phys. Rev. Lett.},\ }\textbf {\bibinfo {volume} {102}},\ \bibinfo {pages}
  {047202} (\bibinfo {year} {2009})}\BibitemShut {NoStop}%
\bibitem [{\citenamefont {Ham}\ \emph {et~al.}(1969)\citenamefont {Ham},
  \citenamefont {Schwarz},\ and\ \citenamefont {O'Brien}}]{Ham69}%
  \BibitemOpen
  \bibfield  {author} {\bibinfo {author} {\bibfnamefont {F.~S.}\ \bibnamefont
  {Ham}}, \bibinfo {author} {\bibfnamefont {W.~M.}\ \bibnamefont {Schwarz}}, \
  and\ \bibinfo {author} {\bibfnamefont {M.~C.~M.}\ \bibnamefont {O'Brien}},\
  }\Doi {10.1103/PhysRev.185.548} {\bibfield  {journal} {\bibinfo  {journal}
  {Phys. Rev.},\ }\textbf {\bibinfo {volume} {185}},\ \bibinfo {pages} {548}
  (\bibinfo {year} {1969})}\BibitemShut {NoStop}%
\bibitem [{\citenamefont {Huang}\ \emph {et~al.}(2004)\citenamefont {Huang},
  \citenamefont {Chang}, \citenamefont {Jeng}, \citenamefont {Guo},
  \citenamefont {Lin}, \citenamefont {Wu}, \citenamefont {Ku}, \citenamefont
  {Fujimori}, \citenamefont {Takahashi},\ and\ \citenamefont {Chen}}]{huang04}%
  \BibitemOpen
  \bibfield  {author} {\bibinfo {author} {\bibfnamefont {D.~J.}\ \bibnamefont
  {Huang}}, \bibinfo {author} {\bibfnamefont {C.~F.}\ \bibnamefont {Chang}},
  \bibinfo {author} {\bibfnamefont {H.-T.}\ \bibnamefont {Jeng}}, \bibinfo
  {author} {\bibfnamefont {G.~Y.}\ \bibnamefont {Guo}}, \bibinfo {author}
  {\bibfnamefont {H.-J.}\ \bibnamefont {Lin}}, \bibinfo {author} {\bibfnamefont
  {W.~B.}\ \bibnamefont {Wu}}, \bibinfo {author} {\bibfnamefont {H.~C.}\
  \bibnamefont {Ku}}, \bibinfo {author} {\bibfnamefont {A.}~\bibnamefont
  {Fujimori}}, \bibinfo {author} {\bibfnamefont {Y.}~\bibnamefont {Takahashi}},
  \ and\ \bibinfo {author} {\bibfnamefont {C.~T.}\ \bibnamefont {Chen}},\ }\Doi
  {10.1103/PhysRevLett.93.077204} {\bibfield  {journal} {\bibinfo  {journal}
  {Phys. Rev. Lett.},\ }\textbf {\bibinfo {volume} {93}},\ \bibinfo {pages}
  {077204} (\bibinfo {year} {2004})}\BibitemShut {NoStop}%
\bibitem [{\citenamefont {Goering}\ \emph
  {et~al.}(2006){\natexlab{a}}\citenamefont {Goering}, \citenamefont {Gold},
  \citenamefont {Lafkioti},\ and\ \citenamefont {Sch\"{u}tz}}]{goering06}%
  \BibitemOpen
  \bibfield  {author} {\bibinfo {author} {\bibfnamefont {E.}~\bibnamefont
  {Goering}}, \bibinfo {author} {\bibfnamefont {S.}~\bibnamefont {Gold}},
  \bibinfo {author} {\bibfnamefont {M.}~\bibnamefont {Lafkioti}}, \ and\
  \bibinfo {author} {\bibfnamefont {G.}~\bibnamefont {Sch\"{u}tz}},\ }\Doi
  {10.1209/epl/i2005-10359-8} {\bibfield  {journal} {\bibinfo  {journal}
  {Europhys. Lett.},\ }\textbf {\bibinfo {volume} {73}},\ \bibinfo {pages} {97}
  (\bibinfo {year} {2006}{\natexlab{a}})}\BibitemShut {NoStop}%
\bibitem [{\citenamefont {Goering}\ \emph
  {et~al.}(2006){\natexlab{b}}\citenamefont {Goering}, \citenamefont
  {Lafkioti},\ and\ \citenamefont {Gold}}]{goering06_comment}%
  \BibitemOpen
  \bibfield  {author} {\bibinfo {author} {\bibfnamefont {E.}~\bibnamefont
  {Goering}}, \bibinfo {author} {\bibfnamefont {M.}~\bibnamefont {Lafkioti}}, \
  and\ \bibinfo {author} {\bibfnamefont {S.}~\bibnamefont {Gold}},\ }\Doi
  {10.1103/PhysRevLett.96.039701} {\bibfield  {journal} {\bibinfo  {journal}
  {Phys. Rev. Lett.},\ }\textbf {\bibinfo {volume} {96}},\ \bibinfo {pages}
  {039701} (\bibinfo {year} {2006}{\natexlab{b}})}\BibitemShut {NoStop}%
\bibitem [{\citenamefont {Huang}\ \emph {et~al.}(2006)\citenamefont {Huang},
  \citenamefont {Lin},\ and\ \citenamefont {Chen}}]{huang06_reply}%
  \BibitemOpen
  \bibfield  {author} {\bibinfo {author} {\bibfnamefont {D.~J.}\ \bibnamefont
  {Huang}}, \bibinfo {author} {\bibfnamefont {H.-J.}\ \bibnamefont {Lin}}, \
  and\ \bibinfo {author} {\bibfnamefont {C.~T.}\ \bibnamefont {Chen}},\ }\Doi
  {10.1103/PhysRevLett.96.039702} {\bibfield  {journal} {\bibinfo  {journal}
  {Phys. Rev. Lett.},\ }\textbf {\bibinfo {volume} {96}},\ \bibinfo {pages}
  {039702} (\bibinfo {year} {2006})}\BibitemShut {NoStop}%
\end{thebibliography}
\end{document}